\pdfoutput=1

\documentclass[a4paper,twocolumn]{article}

\usepackage{amsmath}
\usepackage{amsfonts}

\usepackage[separate-uncertainty=true]{siunitx}
\DeclareSIUnit{\fps}{fps}
\usepackage{subcaption}
\usepackage{comment}
\usepackage{cases}  

\usepackage{mathtools}

\usepackage{siunitx}
\DeclareSIUnit\angstrom{\text {Å}}
\usepackage{float}
\usepackage{graphicx} 

\usepackage{hyperref}

\hypersetup{
    pdftitle={Single-exposure X-ray dark-field imaging via a dual-energy propagation-based setup},    
    pdfauthor={Jannis Ahlers},     
    pdfsubject={Single-exposure X-ray dark-field imaging via a dual-energy propagation-based setup},   
    pdfcreator={Jannis Ahlers},   
    pdfproducer={Jannis Ahlers}, 
    pdfkeywords={X-ray} {Imaging} {Dark-field} {Harmonics} {Dual-energy} {Propagation-based imaging} {PBI}, 
}

\usepackage[capitalise]{cleveref}
\creflabelformat{equation}{#2\textup{#1}#3}

\usepackage[margin={2cm,2cm}]{geometry}
\title{Single-exposure X-ray dark-field imaging via a dual-energy propagation-based setup}
\author{Jannis N. Ahlers$^{1,*}$, Konstantin M. Pavlov$^{2,1,3}$, Marcus J. Kitchen$^{1}$, Stephanie A. Harker$^{1}$,\\Emily J. Pryor$^{1}$, James A. Pollock$^{1}$, Michelle K. Croughan$^{1}$,Ying Ying How$^{1}$,\\Marie-Christine Zdora$^{1}$, Lucy F. Costello$^{1}$, Dylan W. O’Connell$^{1}$,\\Christopher Hall$^{4}$, and Kaye S. Morgan$^{1}$}
\date{\small{$^{1}$ School of Physics and Astronomy, Monash University, Clayton VIC 3800, Australia\\
$^{2}$ School of Physical and Chemical Sciences, University of Canterbury, Christchurch 8140, New Zealand\\
$^{3}$ School of Science and Technology, University of New England, Armidale NSW 2351, Australia\\
$^{4}$ Australian Synchrotron, ANSTO, Clayton VIC 3168, Australia\\
$^{*}$ Corresponding author: jannis.ahlers@monash.edu}}

\begin{document}

\twocolumn[
  \begin{@twocolumnfalse}
    \maketitle
    \begin{abstract}
        \noindent
        X-ray dark-field imaging visualises scattering from sample microstructure, and has found application in medical and security contexts. While most X-ray dark-field imaging techniques rely on masks, gratings, or crystals, recent work on the Fokker--Planck model of diffusive imaging has enabled dark-field imaging in the propagation-based geometry. Images captured at multiple propagation distances or X-ray energies can be used to reconstruct dark-field from propagation-based images but have previously required multiple exposures. Here, we show single-exposure dark-field imaging by exploiting the harmonic content in a monochromatised synchrotron beam and utilising an energy-discriminating photon-counting detector to capture dual-energy propagation-based images. The method is validated by filming time-varying samples, showing the advantage of the dark-field contrast in analysing dynamic evolution. We measure and adjust for the impact of detector charge-sharing on the images. This work opens the way for low-dose and dynamic dark-field X-ray imaging without the need for a high-stability set-up and precision optics.
    \end{abstract}
    \bigskip
  \end{@twocolumnfalse}
]


\section{Introduction}

Diffusive dark-field imaging is a burgeoning X-ray imaging modality that provides contrast from regions of the sample containing random sub-resolution microstructure that diffusely scatters the beam. It has promising applications in medical \cite{urban:2022:qqaeudcr}, security \cite{miller:2013:pcxissa}, and industrial \cite{endrizzi:2015:exdivdcs} contexts. As many applications require dynamic or high-throughput imaging, a key aim of recent work in dark-field imaging has been minimising the number of exposures required to reconstruct a dark-field image.     

Several methods have been developed for X-ray dark-field imaging. In the case of imaging with minimal exposures, most of these methods are based on structured illumination; they pattern the X-ray wavefield using a grid, grating, or random mask and then analyse how that pattern is modified by the introduction of the sample. When the pattern is resolved by the detector, a dark-field reconstruction can be made from a single exposure of the sample \cite{wen:2010:sxdpcdiuttg,kagias:2016:2hssss} via comparisons with a reference image acquired prior.    
Recent work has shown that diffusive dark-field retrieval is also possible from propagation-based images, without needing a reference image \cite{leatham:2023:xdprofpe,ahlers:2024:xdspi}. In near-field conditions, the intensity $I(x,y,z)$ of a coherent monochromatic wavefield after propagating a distance $z=\Delta$ from the exit-surface of the sample at $z=0$ can be modelled using the X-ray Fokker--Planck equation \cite{paganin:2019:xfpepi,morgan:2019:afpegxpdi}:
\begin{equation}
    \label{eq:FP}
    I_{z=\Delta} \approx I_{z=0} - \frac{\Delta}{k} \nabla_\perp \cdot [I\nabla_\perp \phi]_{z=0} + \Delta^2 \nabla_\perp^2[D I]_{z=0},  
\end{equation}
where $k = 2\pi / \lambda$ is the wavenumber, $\lambda$ is the wavelength, $\nabla_\perp \equiv (\partial_x, \partial_y)$ is the transverse gradient operator, $\phi(x,y,z)$ is the phase-shift of the wavefield, and $D(x,y)$ is the X-ray Fokker--Planck diffusion coefficient. Dark-field retrieval consists of solving this equation for the diffusion coefficient. Under the assumption of a single-material sample, attenuation and phase-shift are linked by the projected sample thickness $T(x,y)$, reducing the problem to two unknowns, $T$ and $D$. This problem can be solved using images taken at two propagation distances \cite{leatham:2023:xdprofpe}. 

Alternatively, let us assume the energy dependence of $D$ is known, and that it can be decomposed as $D = D_0 \Gamma(\lambda)$ where $D_0$ is independent of energy and $\Gamma(\lambda)$ is some function that encodes the energy dependence. In this case, two images taken at different energies can be used to reconstruct the projected thickness using spectral propagation-based dark-field imaging (SPB-DF) \cite{ahlers:2024:xdspi}. Let the two subscripts $1$ and $2$ denote the two energies. The projected thickness $T$ can be calculated as \cite{ahlers:2024:xdspi}: 
\begin{equation}
\label{eq:T}
\begin{split}
    T &= \mathfrak{F}^{-1} \left [ \frac{\mathfrak{F} \left[ \mu_2 \Gamma(\lambda_2) I_1 - \mu_1 \Gamma(\lambda_1) I_2 \right]}
    {f(\lambda_1,\lambda_2,\vec{\xi})} \right ]
    \\ &+ \frac{\mu_1 \Gamma(\lambda_1) - \mu_2 \Gamma(\lambda_2)}{\mu_1 \mu_2 (\Gamma(\lambda_1) - \Gamma(\lambda_2))},
\end{split}
\end{equation}
where
\begin{equation}
    \begin{split}
        f(\lambda_1,\lambda_2,\vec{\xi}) &= \mu_1 \mu_2 (\Gamma(\lambda_1) - \Gamma(\lambda_2)) \\ &- 4 \pi^2 |\vec{\xi}|^2 \Delta (\delta_1 \mu_2 \Gamma(\lambda_2) - \delta_2 \mu_1 \Gamma(\lambda_1)),
    \end{split}
\end{equation}
and $\mu$ is the linear attenuation coefficient, $\delta$ is the real refractive index decrement, and $\mathfrak{F}[f(x,y)](\vec{\xi}) = \iint_{-\infty}^{\infty} f(x,y) \exp{(-2\pi i \vec{r} \cdot \vec{\xi})} \mathop{d\vec{r}}$ is the 2D Fourier transform. To correct for an assumption of weak attenuation, $T$ is iteratively reconstructed using \cref{eq:T}. Having reconstructed $T$, the only unknown left in the Fokker--Planck equation is the diffusion coefficient $D$. There are two possible methods to retrieve $D$: a global solution found with a Poisson solver \cite{leatham:2023:xdprofpe,paganin:2023:pdr,ahlers:2024:xdspi}; or a local solution based on comparing local visibility in the flat-field-corrected image $V_{I_1}$ with that in a dark-field-free image $V_{I_\text{DF-free}}$ estimated by numerically propagating the wavefield from the sample to the downstream detector, with the phase and intensity determined from the reconstructed sample thickness \cite{morgan:2019:afpegxpdi, ahlers:2024:xdspi}: 
\begin{numcases}{D=}
    \frac{e^{\mu T}}{\Delta^2} \nabla_\perp^{-2} \biggl[ I - \underbrace{\left(1 - \frac{\delta \Delta}{\mu} \nabla^2_\perp \right) e^{-\mu T}}_{I_{\text{DF-free}}} \biggr] & \text{Global}
    \\
    \frac{-p^2}{4 \pi^2 \Delta^2} \ln{\left( \frac{V_{I_1}}{V_{I_\text{DF-free}}} \right)}, & \text{Local}
\end{numcases}
where $\nabla_\perp^{-2}$ is a numerical implementation of an inverse Laplacian with a fine-tuned Tikhonov regularisation parameter $\varepsilon$ \cite{leatham:2023:xdprofpe, ahlers:2024:xdspi}, $p$ is the period of any local image texture, and $V$ denotes visibility, measured locally around each pixel as the ratio of the standard deviation to the mean \cite{ahlers:2024:xdspi}. Which method is preferable depends on various factors; the local reconstruction can be more stable but requires relatively high-frequency local contrast to be generated by the sample, while the global reconstruction can give higher spatial resolution but requires strong dark-field gradients \cite{ahlers:2024:xdspi}. A more thorough examination of these methods is given in section~1 of the supplemental document.    

Dual-energy X-ray imaging (DEXI) is routinely used in clinical imaging. Most technologies for DEXI rely on either two separate sources and detectors (typically oriented at right angles) or rapid switching between two energies using, for example, kV-switching or a rotating filter. Single-exposure DEXI requires a dichromatic source and a detector setup capable of simultaneous and separate imaging of the two energies. A dichromatic source can be realised by the exploitation of higher-harmonic radiation from a crystal monochromator, which can be combined with a filtered dual-phosphor \cite{carnibella:2012:sdxiuss} or photon-counting detector \cite{bisogni:2002:dsbbxsscdsdma} to achieve single-exposure DEXI.       

We report here the experimental implementation of dual-energy X-ray imaging at a synchrotron of a first and third monochromator harmonic using a photon-counting detector and its use for single-shot dark-field imaging. This is the first time reference- and optics-free single-shot X-ray dark-field imaging has been achieved.     

\section{Method}
\label{sec:method}

Imaging was conducted at the Imaging and Medical Beamline (IMBL) of the Australian Synchrotron. The IMBL consists of a superconducting multi-pole wiggler at \qty{1.4}{\tesla}, which produces a broad spectrum that is filtered by a bent double-Laue crystal monochromator \cite{stevenson:2017:qcxbasimbi}. The samples were placed \qty{4}{\meter} upstream of the detector in hutch~3B, approximately \qty{140}{\meter} away from the source. 

The detector was a photon-counting EIGER2~CdTe~3M\nobreakdashes-W (DECTRIS~AG, Switzerland) with a \qty{75}{\um} pixel size, in its two-threshold mode. Note that this detector does not include any charge-sharing correction.

Each detector pixel counts all photons above the two set thresholds, recorded as two images per exposure. We subtracted the upper threshold counts from the lower threshold counts to isolate the number of counts into two energy bins, which we henceforth refer to as Bin~A (between the two thresholds, the lower energy bin) and Bin~B (above the upper threshold).

\begin{figure}
    \centering
    \includegraphics[width=\columnwidth]{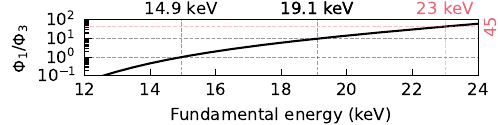}
    \caption{Calculated ratio of fundamental to third-harmonic detected flux for different monochromator settings, suggesting equal flux at \qty{14.9}{\keV} and a $10 \mathbin{:} 1$ ratio at \qty{19.1}{\keV}. Empirically, \qty{23}{\keV} was found to give an approximately $10 \mathbin{:} 1$ ratio of counts in Bins A and B.}
    \label{fig:fluxcalc}
\end{figure}

A key requirement for dual-energy imaging is balancing the flux at each energy to avoid unnecessary dose or detector saturation. A calculation of the ratio of the fundamental to third harmonic flux $\Phi_1/\Phi_3$ at different monochromator settings was carried out, taking into account the source spectrum, the various beamline elements (such as filters), the detector quantum efficiency, and assuming \qty{100}{\percent} monochromator throughput at both the fundamental and third harmonic. The results are shown in \cref{fig:fluxcalc}, suggesting an optimal monochromator setting giving equal flux at \qty{14.9}{\keV}.  
However, this calculation did not consider the escape of fluorescence photons or charge-sharing effects in the detector. Charge sharing occurs when the charge cloud from conversion of a high-energy photon spreads over several pixels. For energies above approximately \qty{30}{\keV}, this leads to spectral efficiency dropping below \qty{20}{\percent} \cite{trueb:2017:asphpcxd}, meaning a large percentage of high-energy photons are erroneously counted as multiple low-energy photons. A balanced flux entering the detector would lead to Bin~A being swamped by third-harmonic events, so the proportion of high-energy flux needs to be reduced. We empirically found a monochromator fundamental energy of \qty{23}{\keV} (with corresponding third-harmonic at \qty{69}{\keV}) to give a ratio of approximately $10 \mathbin{:} 1$ counts in Bins~A and B, decreasing the effect of charge sharing on Bin~A without making Bin~B too noisy. On the Eiger detector, we set the upper threshold to \qty{30}{\keV} to be well above the primary energy \qty{23}{\keV}, taking into account the energy resolution of approximately \qty{2}{\keV} \cite{trueb:2017:asphpcxd}. The lower threshold on the Eiger was set to \qty{4}{\keV} to eliminate electronic noise, but could in future be set up to half of \qty{23}{\keV} to help reduce charge-sharing effects \cite{trueb:2017:asphpcxd}.    

\begin{figure}
    \centering
    \includegraphics[width=\columnwidth]{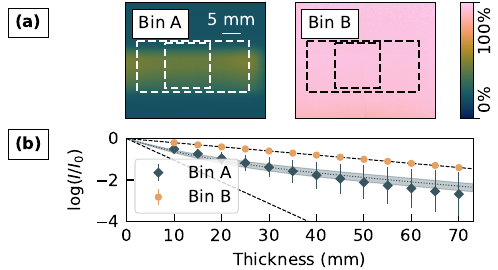}
    {\phantomsubcaption\label{fig:fit-fractions}}
    {\phantomsubcaption\label{fig:fit-line}}
    \caption{Measurement of the energy content of the beam in both bins of the detector by fitting an admixed attenuation model (\cref{eq:admix}). (\subref{fig:fit-fractions}) The resulting fraction of third harmonic counts in each pixel, for each of the two bins. The dashed lines show the FOVs used for the two samples. Relative errors in fitting $f$ do not exceed 11\% in the whole image or 8\% within the larger FOV. (\subref{fig:fit-line}) Attenuation curves found by averaging the pixels in the larger FOV while imaging water-equivalent slabs of the shown thicknesses, fitted to the attenuation model (\cref{eq:admix}). Dashed lines indicate pure \qty{23}{\keV} (steeper) and \qty{69}{\keV} (shallower) attenuation. The fit for Bin~B overlies the \qty{69}{\keV} line exactly and is not shown. Error bars and shading show $3\sigma$.}
    \label{fig:fit}
\end{figure}

To measure the resulting energy content of each bin, we imaged the flat beam with slabs of varying thicknesses of a water equivalent phantom (Solid Water HE, Gammex, USA) inserted. Assuming no energies apart from the fundamental and third harmonic are present, the attenuation of the beam can be modelled using an admixed form of the Beer-Lambert law \cite{tran:2003:qdmssxeshc}: 
\begin{equation}
    \label{eq:admix}
    \frac{I}{I_0} = (1-f)e^{-\mu_1 T} + f e^{-\mu_3 T},
\end{equation}
where the subscripts $1$ and $3$ denote the fundamental and third harmonic, respectively, and the parameter $f$ denotes the fraction of the beam that is the third harmonic energy. Linear attenuation coefficients for the phantom were calculated with \texttt{xraylib} \cite{schoonjans:2011:xlxird}, using a density of \qty{1.032}{\gram\per\cm\cubed} and the mass fractions reported in \cite{schoenfeld:2015:wepm1b}. This model was fit at each pixel for each bin using Nelder-Mead optimisation, applying a bound of $[0,1]$ to $f$. The resulting third-harmonic fractions are shown in \cref{fig:fit-fractions}. Bin~B consists almost solely of third-harmonic. Due to charge-sharing, Bin~A consists of a mixture of energies, with the fraction of third-harmonic varying vertically due to the narrower profile of the higher energy beam. Before reconstruction, images of the two samples were cropped to the two regions-of-interest (ROIs) outlined in \cref{fig:fit-fractions}. A single value for the fraction of the third harmonic $f$ in each bin was found by fitting the model to the mean intensities within the larger ROI [see \cref{fig:fit-line}], giving $f_A = \qty{41 \pm 3}{\percent}$ and $f_B = \qty{99.76 \pm 0.08}{\percent}$. As $f_\text{B}$ is approximately unity, we can treat Bin~B as monochromatic. However, Bin~A comprises a mix of the fundamental and third harmonic energies. To enable dark-field reconstruction using the SPB-DF algorithm, the energy-dependent parameters $\lambda$, $\mu$, and $\delta$ were weighted to calculate an effective equivalent, giving e.g.\ 
$\lambda_1 = (1-f_\text{A}) \lambda_{\qty{23}{\keV}} + f_\text{A} \lambda_{\qty{69}{\keV}}$ \cite{arhatari:2008:piupxls}.

\section{Results}

The described technique enables rapid imaging, only restricted by the exposure time required to achieve sufficiently high photon counts that do not exceed the maximum count rate of the detector. To demonstrate this, two time-varying samples were imaged.
\begin{figure} 
    \centering
    \includegraphics[width=\columnwidth]{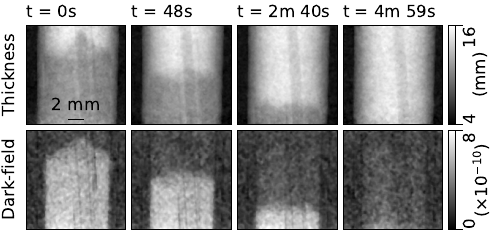}   
    \caption{Water slowly saturates a column of glutinous rice flour, increasing the sample's density but reducing the strong scattering from flour--air interfaces that is captured by the dark-field images. See \textbf{Visualisation 1} for the full time sequence. \label{fig:wetriceflour}}
\end{figure}
The first sample consisted of a plastic test tube filled with glutinous rice flour behind a container holding a thin layer of \qtyrange{250}{300}{\um} diameter polymethyl methacrylate (PMMA) microspheres (\textit{Cospheric LLC}, USA). The sample was imaged as water was added. The flour was first manually agitated to reduce clumping and remove trapped air, and a small piece of plastic tubing was inserted into the flour to facilitate air escaping. As water dripped into the test tube, the sample was imaged with a \qty{1}{\s} exposure time at a rate of \qty{1}{\fps}. 
The projected thickness of the sample (\cref{fig:wetriceflour}, top row) was reconstructed using twenty iterations (\cref{eq:T}), assuming the sample was composed of water, with a dark-field dependence of $\Gamma(\lambda) = \lambda^3$ (see \cite{ahlers:2024:xdspi} for justification). A detailed description of the two methods of reconstructing $D$ and their strengths and weaknesses is given in section~1 of the supplemental document, together with a justification for the use of global versus local reconstruction for the two samples described here. For this sample, the local method was selected to reconstruct the diffusion coefficient (\cref{fig:wetriceflour}, bottom row). The mean period of the sample's texture was measured to be $p = 4.8\thinspace  \text{pixels}$ (method described in section~2 of the supplemental document). As this sample was changing relatively slowly, a $3\times3\times3$ median filter was applied to both signals to improve the signal-to-noise ratio. As the water filters down, the increased density increases the projected thickness, but the change from flour--air interfaces to flour--water interfaces reduces the dark-field scattering signal.   

\begin{figure*}
    \centering
    \includegraphics[width=\textwidth]{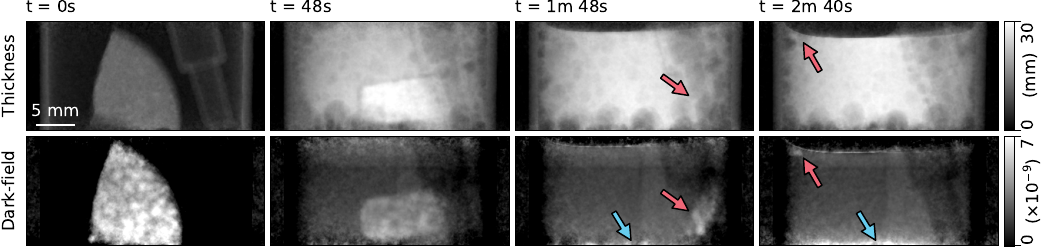}
    \caption{Dissolution of a piece of an effervescent tablet in water. Before the water is introduced, at $t = \qty{0}{\s}$, we see that the solid plastic test tube and syringe do not appear in the dark-field image. When the water is introduced, the reaction with the tablet is rapid, and the tablet quickly shrinks, making it difficult to distinguish from the surrounding solution in the attenuation image; in contrast, the remnant of the tablet can be clearly distinguished in the dark-field image (red arrows). This difference is particularly pronounced in the video, available as \textbf{Visualisation 2}. The build-up of undissolved sediment settling on the bottom can clearly be seen in the dark-field images (blue arrows). \label{fig:barocca}} 
\end{figure*}

The second sample consisted of a piece of an effervescent vitamin tablet (Berocca, Bayer~AG) in a plastic test tube, onto which water was dripped. Imaging and thickness reconstruction were conducted with the same settings as the rice flour sample, except that the sample was assumed to be composed of water. The thickness and global dark-field reconstructions ($\varepsilon = \qty{2.5E-5}{\per\um\squared}$) at different time points are shown in \cref{fig:barocca}. The bulk of the tablet dissolves quickly, leading to vigorous bubbling and an increase in the density of the surrounding solution. A small dark-field signal is also seen from the water/solution; we hypothesise this could be caused by scattering from tiny bubbles or suspended particles. As the tablet shrinks and is pushed around by the bubbles, it becomes challenging to distinguish it from the solution in the projected thickness images. However, even a small remnant of the granular tablet still gives a relatively strong X-ray dark-field signal, and hence it can be better differentiated.               

\section{Discussion}


Propagation-based X-ray dark-field imaging using two distances \cite{leatham:2023:xdprofpe} or monochromator positions \cite{ahlers:2024:xdspi} can suffer from alignment problems caused by slight differences in beam angle or magnification. Such issues are eliminated in a single-shot approach, as shown here. 

The most significant issue in using photon-counting detectors for high-energy spectral measurements is charge-sharing \cite{trueb:2017:asphpcxd}. The combination of (1) a large difference in energy between fundamental and third harmonic and (2) suppression of charge-sharing effects by reducing the high-energy flux was effective in our experiments in ensuring the energy content of each bin was sufficiently different to allow for good qualitative results. However, quantitative accuracy would require using charge-sharing correction, such as Medipix3's charge summing mode \cite{gimenez:2011:scmumsb}.  

The need to reduce the relative high-energy flux to minimise the impact of charge-sharing 
increased the noise in the Bin~B image and necessitated relatively long exposure times of \qty{1}{\s} in our experiments (limiting the frame rate to \qty{1}{\fps}). We emphasise that there is no fundamental restriction on the achievable frame rate using this technique beyond that introduced by the source flux and detector limits, particularly if charge-sharing correction could be used.  

Spectral imaging and photon counting detectors are already prevalent in synchrotron and laboratory contexts, implying our technique could be easily implemented. Using third harmonics means our technique is compatible with beamline optics that accept harmonic energies, such as Fresnel zone plates (FZP).   

\section{Conclusion}

Dark-field X-ray imaging provides full-field measurements of small-angle scattering and has a variety of promising applications. Based on the X-ray Fokker--Planck equation under a single-material assumption, two propagation-based images at different distances or energies can be used to reconstruct the dark-field image. We report using monochromator harmonics with an energy-resolving photon-counting detector to achieve single-exposure dark-field imaging. This removes the need to align multiple images and enables time-resolved scattering sample imaging. 

\subsubsection*{Funding}
National Health and Medical Research Council (APP2011204); Australian Research Council (DP230101327).

\subsubsection*{Acknowledgments}
Jannis Ahlers is supported by an Australian Government Research Training Program (RTP) Scholarship and an AINSE Ltd.\ Postgraduate Research Award (PGRA). The experiments were completed at the Australian Synchrotron, part of ANSTO, under proposals 20670 and 21857, supported by the Australian Research Council (DP230101327) and the National Health and Medical Research Council (APP2011204). We thank Yakov Nesterets for useful discussions.   

\subsubsection*{Disclosures}
\noindent The authors declare no conflicts of interest.

\subsubsection*{Data availability} Data underlying the results presented in this paper are not publicly available at this time but may be obtained from the authors upon reasonable request.

\bibliographystyle{abbrv}
\bibliography{references}

\end{document}


\maketitle

\section{Local and global dark-field reconstruction}

\begin{figure}
    \centering
    \includegraphics[scale=1]{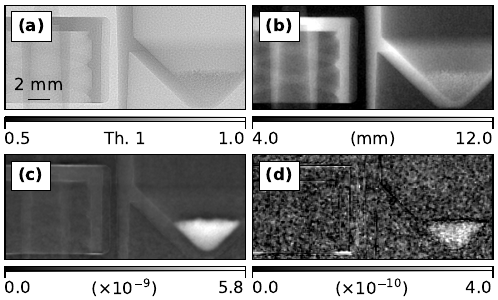}
    {\phantomsubcaption\label{fig:polysty-I_E1}}
    {\phantomsubcaption\label{fig:polysty-T}}
    {\phantomsubcaption\label{fig:polysty-D_glob}}
    {\phantomsubcaption\label{fig:polysty-D_loc}}
    \caption{Single-exposure propagation-based X-ray dark-field imaging of a sample of \qty{1}{\um} polystyrene items, including microspheres in a solid tube. (\subref{fig:polysty-I_E1})~The flat-field corrected image at the lower energy, and reconstructions of (\subref{fig:polysty-T}) the projected thickness and dark-field using (\subref{fig:polysty-D_glob}) the global and (\subref{fig:polysty-D_loc}) the local methods. \label{fig:polysty}}
\end{figure}

\begin{figure}
    \centering
    \includegraphics[scale=1]{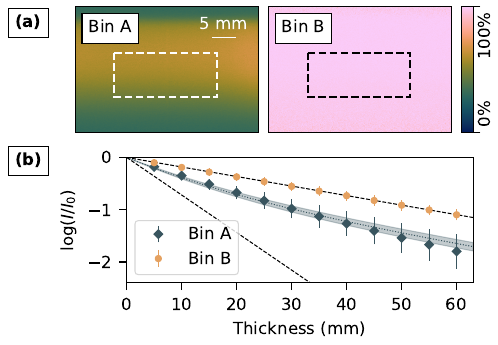}
    {\phantomsubcaption\label{fig:fit-fractions}}
    {\phantomsubcaption\label{fig:fit-line}}
    \caption{Measurement of the energy content of the beam for the static sample. (\subref{fig:fit-fractions}) The resulting mixing fraction in each pixel, with the ROI of the sample. Relative errors in fitting $f$ do not exceed 10\% in the whole image or 3\% within the FOV. (\subref{fig:fit-line}) Attenuation curves found by averaging the pixels in the FOV while imaging water-equivalent slabs of the shown thicknesses, fitted to the admixed attenuation model. Dashed lines indicate pure \qty{27}{\keV} (steeper) and \qty{81}{\keV} (shallower) attenuation curves. The fit for Bin~B overlies the \qty{81}{\keV} line exactly and is not shown. As errors are small, $3\sigma$ are plotted.}
    \label{fig:fit}
\end{figure}

Ahlers et al.\ \cite{ahlers:2024:xdspi}\ presented two methods to reconstruct the diffusion coefficient $D$ after reconstructing the projected thickness. Both methods are based on measuring a difference between the lower energy image $I_1$ (which has stronger dark-field effects than the higher energy image) and $I_\text{DF-free}$, an image created by Fresnel propagating the exit-wavefield calculated using the reconstructed projected thickness. $I_\text{DF-free}$ does not contain dark-field effects, as the reconstructed projected thickness does not include any microstructural information, and the (low-resolution) Fresnel propagation does not explicitly model dark-field effects (unlike, for example, Fokker--Planck propagation). Therefore, any remaining differences between $I_1$ and $I_\text{DF-free}$ theoretically stem from dark-field scattering. 
How these differences are used to reconstruct the diffusion coefficient $D$ varies between the two methods, with the consequence that different samples (and therefore images) may be better suited to one reconstruction method over another. Here, we briefly review the two methods of reconstructing $D$, summarise their behaviour, and explain the choices made in reconstructing the two dynamic samples in the paper.  

The \emph{local} approach is inspired by structured-illumination dark-field imaging, but uses the sample's texture itself as the (self-)reference pattern. The dark-field is measured as a change in visibility of this texture between $I_1$ and $I_\text{DF-free}$, with the visibility being measured in a window around each pixel. As in structured illumination techniques, the size of this window should cover approximately one period of the texture, and the texture should have good visibility in the dark-field free image \cite{zdora:2018:saxspdi}. While the window limits the resulting spatial resolution, the use of a relatively large region around each pixel to quantify the dark-field makes the method robust to image noise \cite{ahlers:2024:xdspi}. On the other hand, strong phase gradients intersecting with the window region can introduce problematic artefacts. Clearly, the local approach requires the sample to produce a high-frequency texture. In addition, the period of this texture should be consistent throughout the sample, although Ahlers et al.\ \cite{ahlers:2024:xdspi}\ suggested that the period and window size used in the calculation of $D$ at each pixel could be varied if the dominant local texture period were first measured.

The \emph{global} approach is based on directly solving the Poisson equation
\begin{equation}
\label{eq:poisson}
    \nabla_\perp^2 \left( D e^{-\mu T} \right) = \frac{1}{\Delta^2}(I_1 - I_\text{DF-free})
\end{equation}
for the diffusion coefficient $D(x,y)$. To do this, we employ a numerical approximation of the inverse Laplacian based on a Tikhonov-regularised Fourier filter \cite{leatham:2023:xdprofpe,ahlers:2024:xdspi}:
\begin{equation}
    \nabla_\perp^{-2} = -\mathfrak{F}^{-1} \frac{1}{4 \pi^2 |\vec{\xi}|^2 + \varepsilon} \mathfrak{F},
\end{equation}
where $\mathfrak{F}$ and $\mathfrak{F}^{-1}$ are the 2D Fourier transform and corresponding inverse transform, $\vec{\xi}$ are the Fourier-space coordinates, and $\varepsilon$ is the Tikhonov regularisation parameter. In section~1 of the supplemental document to \cite{ahlers:2024:xdspi}, it was noted that, in the absence of strong and consistent local sample contrast, the main contribution to dark-field contrast seems to come from a term proportional to $\nabla_\perp^2 D(x,y)$. This effect is also explained in \cite{morgan:2019:afpegxpdi}, particularly figure~4 and the associated text. The result is that solving the Poisson equation \cref{eq:poisson} requires strong variations in $D$ from small or strongly scattering dark-field-producing parts of the sample.

In summary, the local approach is appropriate for samples with high-visibility, high-frequency texture (which does not introduce strong phase fringes), but where the dark-field signal is expected to be relatively smooth and a high spatial resolution is not needed to assess the reconstructed dark-field images. This is exactly the case in the rice flour sample. The significant feature of interest, the advancing boundary between saturated and unsaturated flour, does not require a high spatial resolution to be identified. On the other hand, the smoothness of the dark-field signal from the very homogenous rice flour distribution makes a global reconstruction difficult. The sample of the dissolving vitamin tablet presents a very different case. Here, there is no consistent local texture, and the strong bubbling in the water creates many strong phase fringes. Many small features of the dark-field structure of the sample are of interest, from the variation within the tablet itself to the settling sediment at the bottom of the test tube. A local reconstruction would, therefore, be inappropriate. However, these features' strong and spatially varying dark-field signal makes the tablet sample an ideal candidate for global dark-field reconstruction.

The two dynamic samples in the main text of the paper exemplify two cases where one approach is significantly better than the other. In order to demonstrate a case where both global and local reconstruction can provide useful results, a static sample with both a strong dark-field signal (and, therefore, dark-field variations) and a local texture was imaged using our single-exposure technique. The sample consisted of two similarly attenuating objects with differing dark-field signals placed side-by-side (see \cref{fig:polysty-I_E1}) behind a screen of \qtyrange{250}{300}{\um} polymethyl methacrylate (PMMA) microspheres. On the right is a plastic test tube containing polystyrene microspheres with a mean diameter of \qty{1.48}{\um} (Corpuscular Inc, USA), and on the left is a solid plastic male Luer lock connector. A slightly different experimental setup was used for this sample than the dynamic samples in the main paper. The monochromator energy was set to \qty{27}{\keV}, and additional filtration was added to the beam, consisting of a pair of in-vacuo aluminium filters (total effective thickness of \qty{2.828}{\mm}) and \qty{20}{\mm} of aluminium filtration ex-vacuo in the imaging hutch, just downstream of the beryllium window. Repeating the measurement of the beam's energy content gave mixing fractions $f_\text{A} = \qty{56 \pm 2}{\percent}$ and $f_\text{B} = \qty{99.3 \pm 0.2}{\percent}$. The exposure time per frame was \qty{20}{\s}, and thirty sample and flat-field frames were recorded and then averaged over. 

In the reconstruction, we used twenty iterations and assumed the sample was composed of PMMA and that the dark-field dependence was $\Gamma \propto \lambda^3$. The resulting projected thickness reconstruction can be seen in (\cref{fig:polysty-T}). The dark-field images were reconstructed using both the global (\cref{fig:polysty-D_glob}) method with $\varepsilon = \qty{3.2E-5}{\per\um\squared}$, as well as the local (\cref{fig:polysty-D_loc}) method with the same period as the rice flour sample. The results are similar to those seen in Figure~9 of Ahlers~et~al.~\cite{ahlers:2024:xdspi}. As expected, the local reconstruction is much noisier. In both local and global reconstruction, the microspheres are significantly more prominent than any other part of the sample. However, unlike in Figure~9(d) of Ahlers~et~al.~\cite{ahlers:2024:xdspi}, some structure of the solid plastic parts of the sample can still be seen in the global reconstruction. We hypothesise that this is related to the usage of energy-weighted effective parameters $\lambda$, $\mu$, and $\delta$ for the lower energy bin.   

\section{Measuring size of sample texture}

\begin{figure}
    \centering
    \includegraphics[scale=1]{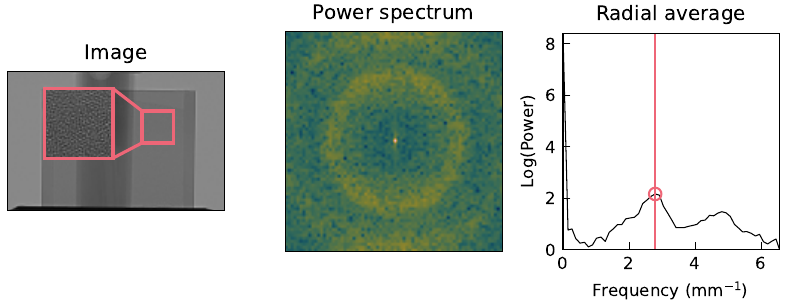}
    \caption{Measurement of the period of rice flour sample texture. The uncropped first frame of the rice flour sample is shown on the left, with a part of the texture with minimal other structure selected for the size measurement. The 2D power spectrum of this section is calculated and shown in the centre. The dominant frequency is found as the first non-zero peak in the radially averaged 2D power spectrum.}
    \label{fig:sizing}
\end{figure}

The local method to reconstruct the Fokker--Planck diffusion coefficient relies on the sample having local features with a consistent size throughout the sample. To maximise spatial resolution, this texture should have a period no larger than a few pixels, mirroring the conditions required in speckle-based dark-field imaging \cite{zdora:2018:saxspdi}. As the period $p$ of the texture is needed for the conversion from visibility to diffusion coefficient, we need to measure the mean period of the sample texture when using the local method. To measure the period of the texture in the rice flour sample, we measured the peak of a radially averaged power spectrum of the texture. This process is illustrated in \cref{fig:sizing}.

\bibliographystyle{plain}
\bibliography{references}